\documentclass[conference]{IEEEtran}
\usepackage[cmex10]{amsmath}
\usepackage{amssymb}
\usepackage{url}
\usepackage{array}
\usepackage{graphicx}
\usepackage{mdwmath}
\usepackage{mdwtab}
\usepackage{eqparbox}
\usepackage{subfig}
\usepackage{color}
\usepackage{stfloats}

\begin{document}

\title{Stay Awhile and Listen: User Interactions in a Crowdsourced Platform Offering Emotional Support}

\author{
    \IEEEauthorblockN{Derek Doran\IEEEauthorrefmark{1}, Samir Yelne\IEEEauthorrefmark{1}, 
    Luisa Massari\IEEEauthorrefmark{2}, Maria-Carla Calzarossa\IEEEauthorrefmark{2}, 
    LaTrelle Jackson\IEEEauthorrefmark{3}, and Glen Moriarty\IEEEauthorrefmark{4}}
    \IEEEauthorblockA{\IEEEauthorrefmark{1}Dept. of Computer Science \& Engineering, Kno.e.sis Research Center, Wright State University, Dayton, OH, USA
    }
    \IEEEauthorblockA{\IEEEauthorrefmark{2}Dept. of Electrical, Computer, and Biomedical Engineering,
    University of Pavia, Pavia, Italy
    }
    \IEEEauthorblockA{\IEEEauthorrefmark{3}School of Professional Psychology, Wright State University, Dayton, OH, USA; 	\IEEEauthorrefmark{4}7 Cups of Tea, Inc.
    }
    \IEEEauthorblockA{Email: \{derek.doran,yelne.2,latrelle.jackson\}@wright.edu, \{luisa.massari,mcc\}@unipv.it, glen.moriarty@7cupsoftea.com }
}

\maketitle

\begin{abstract}
Internet and online-based social systems are rising as the dominant mode of communication in society. 
However, the public or semi-private environment under which most online communications operate under
do not make them suitable channels for speaking with others about personal or emotional problems. 
This has led to the emergence of online platforms for emotional support offering free, anonymous, and confidential 
conversations with live listeners. Yet very little is known about the way these platforms are utilized, and if their
features and design foster strong user engagement. This paper  
explores the utilization and the interaction features of hundreds of thousands of users on 
{\em 7 Cups of Tea}, a leading online platform offering online emotional support. It dissects the user's
activity levels, the patterns by which they engage in conversation with each other, and uses machine learning methods to find factors promoting engagement. The study may be the first to measure activities and interactions in a large-scale online social system that fosters peer-to-peer emotional support. \end{abstract}
\IEEEpeerreviewmaketitle

\section{Introduction and Motivation}
Internet and online-based social platforms encompassing
online social networks such as Facebook, LinkedIn, and Twitter
and messaging services like Snapchat and Kik
are rising as the dominant way people in society communicate with each other. 
On these platforms, users are surrounded by `friends' or `colleagues' who may
happy to help a person presently going through a period of emotional distress. Yet
the public or semi-public nature of these platforms as well as the permanency of their communication records
mean they are less than ideal mediums to seek and receive emotional support. 
There is therefore a need 
for online social systems that offer private, anonymous, quick, and live emotional support
for those who prefer to communicate online and need immediate help~\cite{binik97,huang96}. 
Existing systems for this purpose vary in regards to the type of support offered, from generic advice for common emotional
conditions\footnote{\url{http://www.stress.org/emotional-and-social-support}}, to offering self-diagnosis for a condition~\cite{zuckerman03}.
Some systems also offer access to a live therapist when a user is suffering from a specific condition, such as 
suicide contemplation\footnote{\url{http://www.crisischat.org}, \url{http://www.befrienders.org}} or after receiving a critical
health prognoses\footnote{\url{http://www.cancersupportcommunity.org}}~\cite{bar08,rochlen04,hemmati14}. Past 
studies of online systems connecting users to a live listener confirm their effectiveness~\cite{barak07}, however they are limited 
to only helping those that suffer from a particular ailment.

\begin{figure}[!t]
\centering
\includegraphics[width=3.5in]{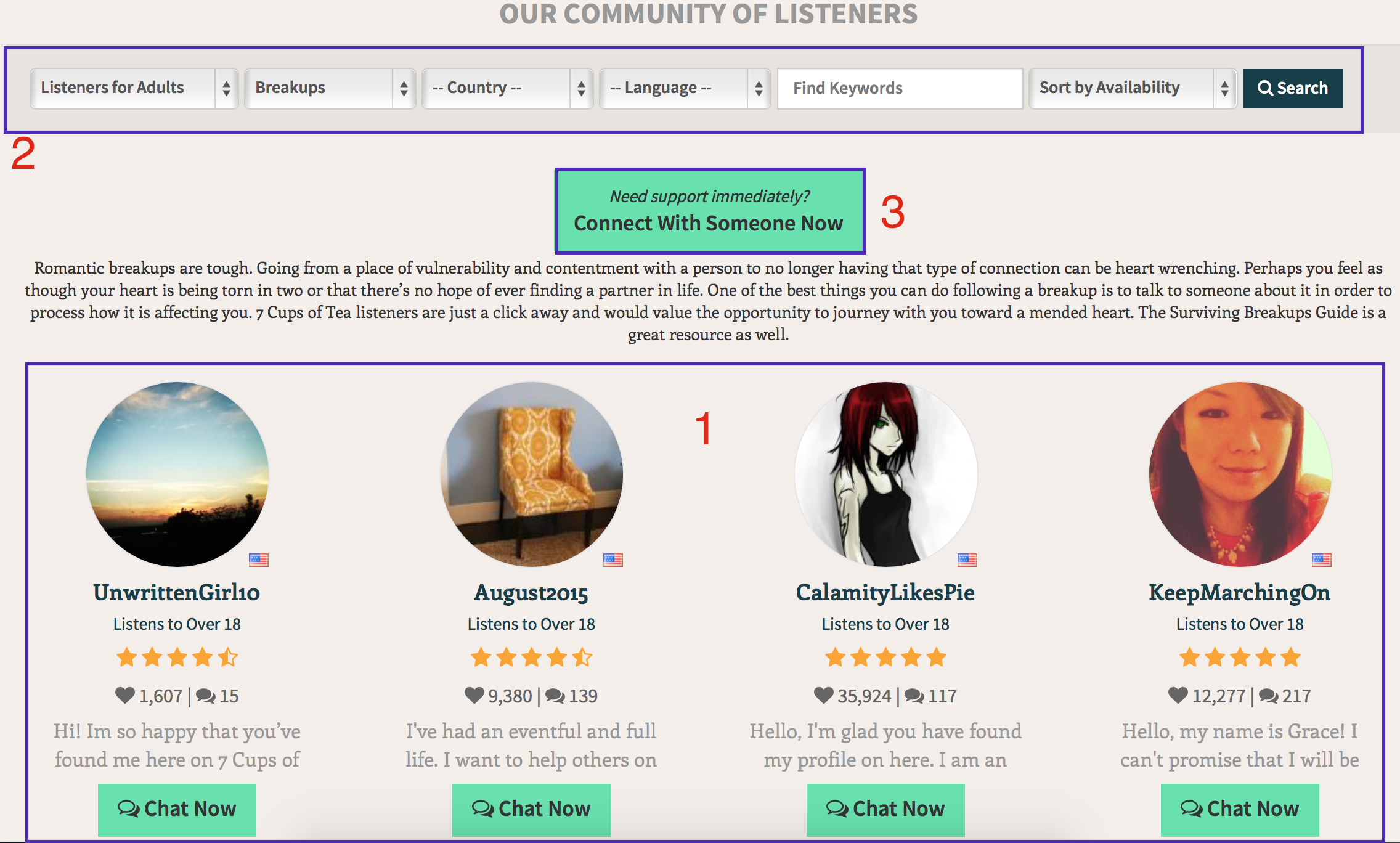}
\caption{Browsing for new listeners in 7cot. (1) Listener profiles; (2) Searching for listener by various criteria; (3) Connecting with a listener immediately.}
\vspace{-15px}
\label{fig:list}
\end{figure}

In order to provide a safe, anonymous space for users to find emotional support for problems of any size,
the online social system {\em 7 Cups of Tea}\footnote{\url{http://www.7cupsoftea.com}} (7cot) was developed. 
As seen in Figure~\ref{fig:list}, 7cot 
fosters an active community or crowd of ``listeners'' who are individuals trained to support people facing a wide range of emotional problems. People needing emotional support may use the service to immediately and privately engage in one-on-one conversations with
listeners or connect to themed group chat rooms. In less than two years, 7cot has attracted a community of hundreds of thousands of members and fostered millions of one-on-one conversations. Its rapid growth suggests a significant demand for 
creating online spaces where users can find and offer emotional support. 

Beyond our knowledge that therapeutic support can be effectively delivered
online~\cite{barak07}, we know very little about 
how online emotional support platforms are utilized by users, the mechanisms with which users connect to listeners, and the design choices that encourage long-term user engagement. 
This paper therefore studies the utilization, interactions, and engagement
of users on 7cot. It specifically explores:~(i) the degree to which activities are performed by different types
of users;~(ii) the interaction structure of member-to-listener conversations
and the relationships among members (listeners) connecting with common 
listeners (members); and~(iii) models that identify the
user and platform features encouraging long-term user engagement. 
The findings are connected to useful insights on how to improve existing platforms, to create effective new ones, 
and to better understand how the Internet is currently used as a `crowdsourced emotional support' tool. 

The layout of this paper is as follows: Section~\ref{sec:overview} gives a broad overview of the 7cot platform. Section~\ref{sec:activity} explores the activity of different types of users on 7cot. Section~\ref{sec:interact} studies the structure of interactions (conversations) between members and listeners. The factors that drive member engagement and model that predicts long-term engagement are presented in Section~\ref{sec:engagement}. A summary of our findings and concluding remarks are 
given in Section~\ref{sec:conc}.

\section{Overview of 7 Cups of Tea}
\label{sec:overview}
7cot launched on December 5th 2013. The service is used by three types of
users: {\em members} choosing to register an account in order to speak
with someone, {\em listeners} who register to listen to the problems of others
and are required to take an online training class, and {\em guests} who choose not to 
register but still wish to converse with listeners. Users may take on multiple types;
for example a member that passes the required training class may become a `hybrid' who
is also a listener. Table~\ref{tab:summary} lists that, as of November 18 2014, the
site is populated by 
87,232 members, 33,601 listeners, and 12,038 hybrid users. The members and 
listeners identify themselves as either a teenager or an adult to 
connect with an appropriate listener. 
Once logged in, self help 
guides are available for users wanting to self-diagnose or support themselves. 

\begin{figure}[!t]
\centering
\includegraphics[width=3.5in]{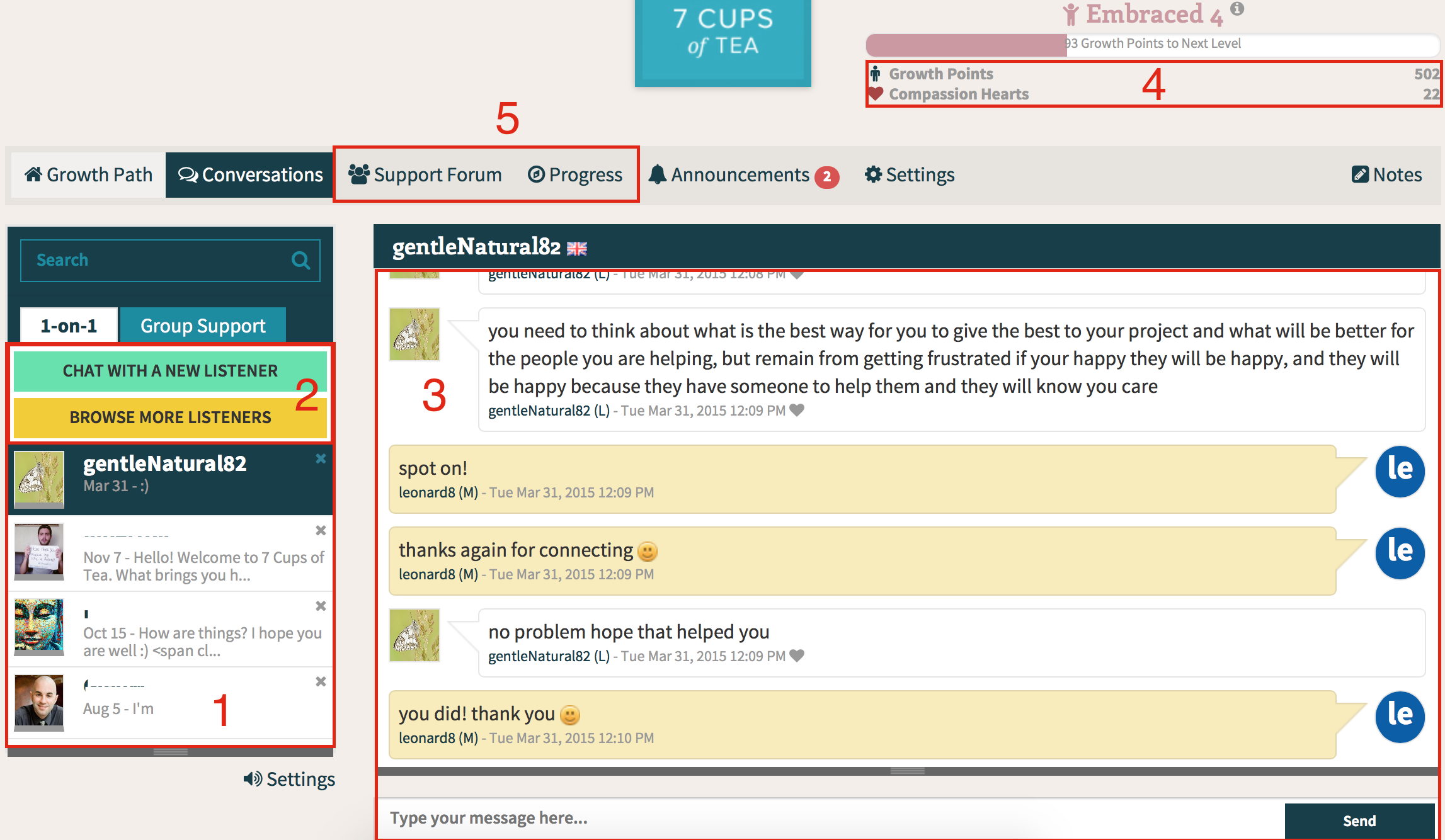}
\caption{7cot member interface. (1) List of current conversations; (2) selections to connect to any listener; (3)
conversation window; (4) progress values; (5) access chat forums and progression metrics.}
\vspace{-15px}
\label{fig:7cot}
\end{figure}

Users communicate with others in three different channels: {\em group chats}, {\em conversations}, or {\em forums}. Group chats are free exchanges that multiple guests, members, and listeners may participate in. Conversations are private exchanges of messages between members or guests and listeners. A conversation is a single, permanent connection lasting for an indefinite amount of time.
Members and guests are able to start a conversation with any listener that is 
currently online, or may search for a listener satisfying some criteria. Members search for listeners through the interface in Figure~\ref{fig:list}. 
It offers a profile of the listeners matching the criteria entered in the top bar, and another option to immediately connect with any active listener should the member be in crisis. The interface members use to access various communication channels is shown 
in Figure~\ref{fig:7cot}. The left panel shows all conversations the user participates in and gives options to 
create new conversations, the right panel is an active conversation,
the top right status bar are values related to members' emotional progress, and the menu options lead to the forum and member profile. 

``Gaming" or ``progress" mechanisms are integrated into the site to represent user reputation and experience. Listeners gradually accrue `cheers' over time, and after attaining certain amounts their `listener level' is upgraded to a more prestigious category. Listeners also achieve `badges' displayed on their profile for accomplishing tasks like helping members facing a specific type of need (e.g., loss of a loved one). Members accrue `growth points' for  performing simple activities such as posting on the forum, or sending messages during a conversation. Accruing enough `growth points' 
will upgrade their `member level', a rank that reflects
a commitment to the site and progress toward improved mental health. 


7cot shared a database capturing the attributes of all users, interactions, and activities performed since its inception on December 5th, 2013 through 
November 18th, 2014. The database includes metadata about every user except for those attributes 
related to the user's true identity and contact information. Attributes of each conversation record were limited to participant
identifiers, the date the conversation commenced, the number 
of messages exchanged by each party, whether the conversation was for a
teenager or adult member, if the conversation was terminated
by the member or listener, and the timestamp of the last message
sent. User behaviors on the site were captured between May 7th 
and November 18th. For privacy reasons, the only actions captured are the number of messages sent, requests made, 
forum posts made, logins,  forum views, help guide views, and page views through the mobile app or Web
browser per user per day. 

\begin{table*}
\begin{center}
\begin{tabular}{r | l || r | l || r | l}
\multicolumn{2}{c}{Participation (a)} & \multicolumn{2}{c}{Actions (Avg. per user per active day) (b) } 
& \multicolumn{2}{c}{Conversations (c) } \\ \hline
Period & Dec 5 - Nov 18 & Period & May 7 - Nov 18 & Period & Dec 5 - Nov 18 \\ 
Num. Conversations & 1.27M & Logins & 2.41 & 
Volume (by Users): & 413,256 (adult); 131,449 (teenager)\\ 

Distinct Forums & 53 & Conversation Messages & 62.28 & 
Volume (by Guests): & 493,365 (adult); 229,918 (teenager)\\

Chatroom Messages & 1.07M & Conversation Requests & 1.83 & 
Type: General & 522,863 (adult); 224,939 (teenager) \\ 
Forum Posts & 82,223 & Forum Posts& 2.93 & 
Type: Personal & 383,758 (adult); 136,428 (teenager) \\ 
Num. Members & 87,232 & Forum Post Views & 6.38 & 
Messages (by Non-Listeners) & 14.77M (adult); 4.28M (teenager)\\
Num. Listeners & 33,601 & Page Views & 15.98 & 
Messages (by Listeners) & 13.54M (adult); 4.12M (teenager)\\
Num. Hybrid  & 12,038 & Help Guide Views& 4.12 &
Terminations & 61,435 (members); 196 (listeners) \\
\end{tabular}
\caption{Summary of 7cot participation, actions, and conversations}
\label{tab:summary}
\end{center}
\vspace{-15px}
\end{table*}

\section{Platform Activity}
\label{sec:activity}
\begin{figure}
\vspace{-15px}
\subfloat[Conversations per month]{\includegraphics[width=1.72in]{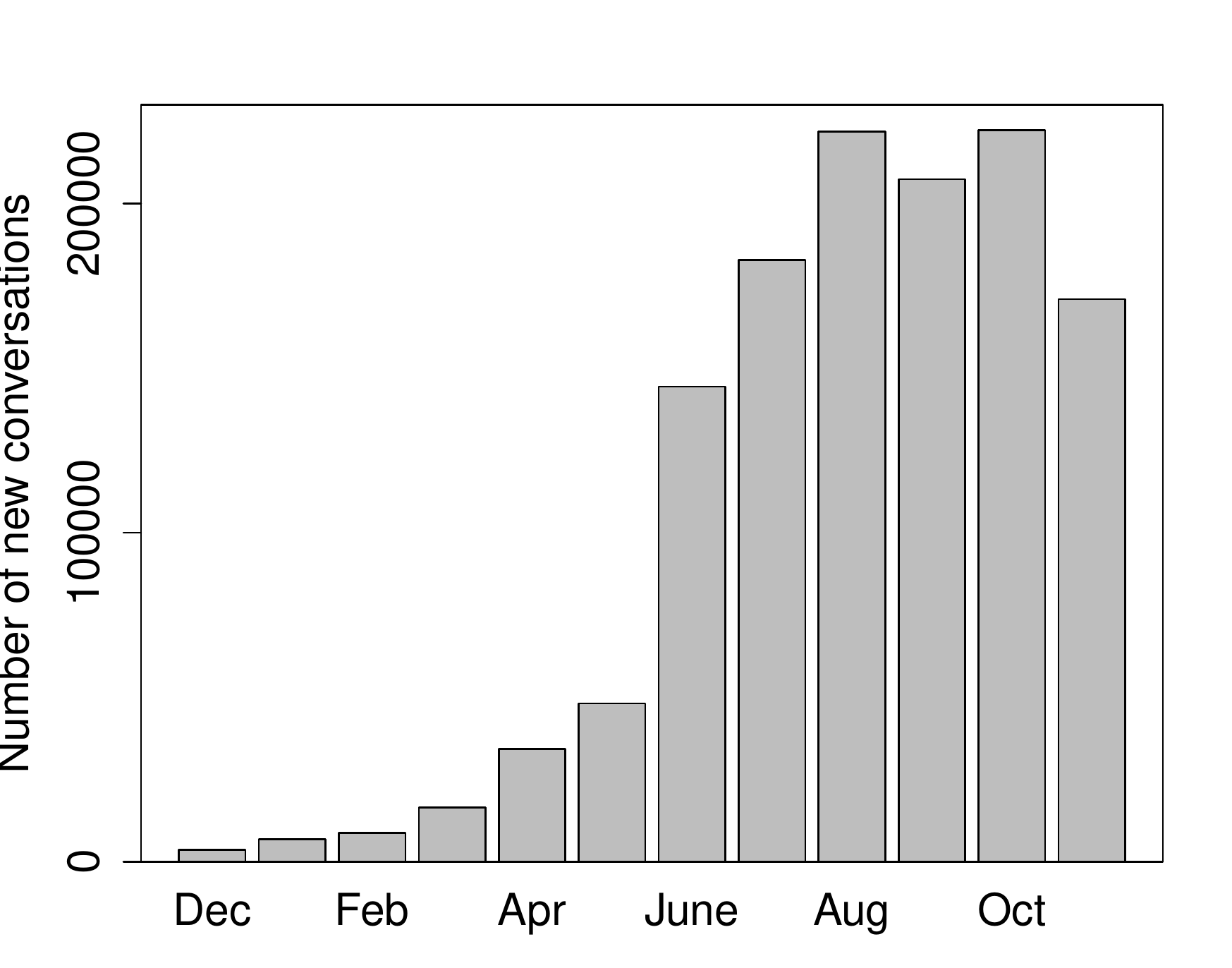}%
	\label{conversations-months}}
\hfill
	\subfloat[Conversations per day of week]{\includegraphics[width=1.72in]{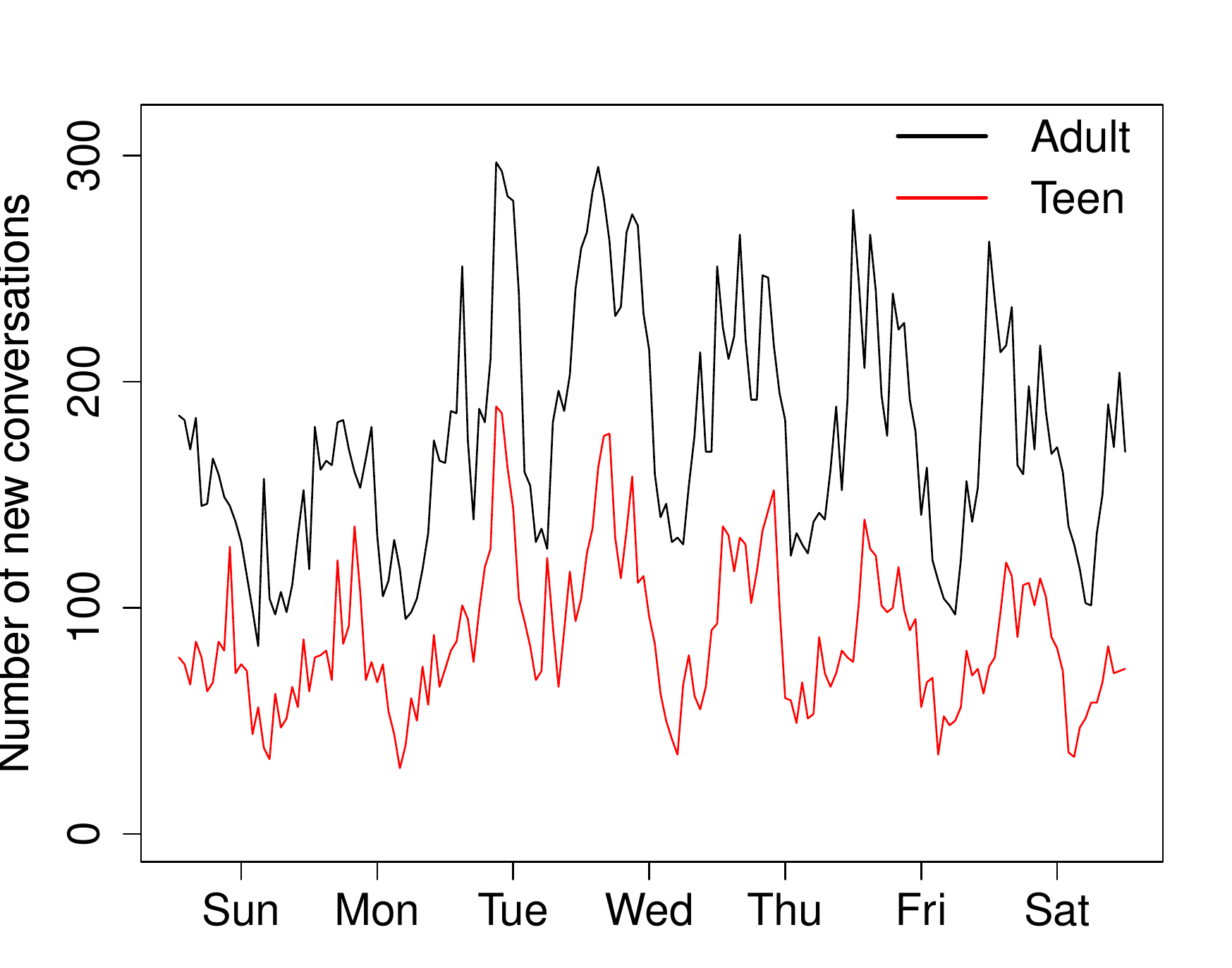}%
	\label{temporal}}
	\caption{Conversation rates on 7 Cups of Tea}
  \vspace{-15px}
\end{figure}

Table~\ref{tab:summary} is divided into three sections that summarize the participation, actions, and conversations held.
The participation statistics in section (a) underscore the size and volume of activity on 7cot. 
In an 11 month span, over 1.27M conversations 
were held between 87,232 members seeking help and 33,601 listeners. 
In addition, 12,038 or 10.0\% of all users are hybrids (both a member and a listener).
The rate at which conversations are initiated 
rose at an exponential pace over 7cot's first 9 months 
as shown in Figure~\ref{conversations-months} (note that conversations
initiated in November 2014 only refer to approximately two weeks).
We also explore the temporal patterns of conversations during 
the week of August 10, 2014 in Figure~\ref{temporal}. The labels on the
$x$ axis are centered to 12pm. The Figure shows a diurnal pattern 
with larger volumes of conversations commencing in the middle weekdays. Furthermore,  
most conversations are initiated in the morning and overnight hours, with a
lull in activity in the midday. These patterns suggest that members have a preference to 
share information during the evening or even overnight hours. 

Section (b) of Table~\ref{tab:summary} 
summarizes the rate of actions undertaken by users per day, 
not counting the days when a user does not perform the action.
For example, users log-in an average of 2.41 times across all days they have logged
in at least once. Furthermore, members connect to an average of 1.83 listeners per 
day they decide to connect to a new listener, and submits an average of 62 messages. 
These statistics indicate that members are not hesitant to reach out with many other
listeners multiple times per day. In fact, 
the platform's ability to let a member communicate with many others, rather 
than a single professional, is a key differentiator between seeking online and offline help. 
For example, 7cot members may listen to the thoughts and perspectives of 
a large number of others, searching for resolution by considering the viewpoints
of many others. Section (b) also shows that forum participation and 
seeking self-guided help are less popular 
compared to participating in one-on-one conversations. 

Section (c) of Table~\ref{tab:summary} lists summary information about  
conversations, broken
down by whether a participant is a teenager or an adult.
Of the 1.27M conversations, more than half are initiated by guests.
This reflects the demand for platforms to let people connect and speak with 
others immediately, without going through an extensive registration process beforehand. 
It also demonstrates an untapped opportunity for a 
platform to transform guests who had positive experiences into members or 
listeners who can further build its community. 
Section (c) also gives the breakdown of conversations that
are ``general" or ``personal". ``General" conversations are ones where a member
asks the platform to connect to any listener, whereas ``personal'' conversations
have a member asking a specific listener to talk to. No matter the type, over
28.5\% of all conversations involve teenagers, 
supporting the hypothesis that young generations find online platforms to be 
a desirable way to express their problems and find support. Users also 
tend to initiate conversations without regard for whom
the listener is, with far more ``general" than ``personal" conversations. 
People seeking emotional support from a crowd may therefore be less interested
in the kind or expertise of a listener. It could also be a reflection of 
7cot's design, which lets members connect to any listener quickly across
many member interfaces. The section also shows 
that approximately 61,631 or 4.9\% of all conversations are 
`canceled' by a user. Canceled conversations are ones where a participant 
decides to permanently terminate a conversation. The relatively
small percentage indicates that users sharing offensive, derogatory, 
or other messages that would lead to conversation termination happens infrequently. 
Users are therefore mostly civil and supportive to each other. 
Conversations are more often terminated by members, 
possibly if they disagree with the listeners suggestions or have
found the conversation unable to solve their emotional problem.  



\section{Interaction Structure}
\label{sec:interact}
We next study the patterns of member engagements with listeners on 7cot. 
The patterns are found through analysis of a network where members and listeners
are connected if they held at least one conversation with each other. 
We also study networks that connect members (listeners) 
to each other if they had a conversation with at least one common listener
(member).  Structural analyses of the networks inform how members are 
choosing to engage with listeners on 7cot, if some subsets of listeners
are more popular than others, and if a pattern of members selectively 
choosing listeners can be seen.


We represent all 7cot interactions as a bipartite network from members to listeners.
We consider all 465,437 conversations that contained 
at least one message sent by either a member or listener (note that guests are excluded
from this analysis and will be the subject of future work). 
Table~\ref{tab:proj} lists the structural features of this bipartite 
network. The network has an average degree $\langle k \rangle$ of $5.39$, 
i.e. members tend to connect to between five or six distinct listeners
during their time on the service. 
This reaffirms the idea that members seek help from a number of others, perhaps
to obtain different viewpoints or thoughts about their emotional problem. 
We also computed
the number of connected components in the network. Only 477 disconnected components exist,
the largest of which (GCC) includes virtually every user (99.2\%) on the platform.
In other words, there are virtually no members or listeners on 7cot who choose to 
exclusively search for and communicate only with each other. 
The single large GCC lets us compute the average path length in the network as
$\bar{d} = \log(|V|/z_1)/\log(z_2/z_1) + 1$, an expression valid for 
networks that are nearly fully connected~\cite{newman}, 
where $z_1$ and $z_2$ are the average number of others a user can reach within
one and two hops respectively. The small average path length $\bar{d} = 3.46$
may be indicative of the existence of a large `core' of members and 
listeners serving as hubs that connect members and listeners to others across
the bipartite structure. 
Listeners in the `core' may thus connect to large and diverse sets of members, i.e., 
are the listeners that connect to members who request to speak with 
any available listener. 

\begin{table}
\setlength\extrarowheight{0.5pt}
\begin{tabular}{r| c | c | c}
 & Bipartite Network & Member Proj. & Listener Proj. \\
 \hline
 $|V|$ & 117,372 & 86,877 & 30,495 \\
 $|E|$ & 465,437 & 12,657,611 & 10,359,604 \\
 $\langle k \rangle$ & 5.39 & 291.39 & 679.43 \\
 $\mathcal{C}$ & N/A & 0.734 & 0.636 \\ 
 $\mathcal{A}$ & N/A & -0.10 & -0.06 \\
 $\bar{d}$ & 3.46 & 2.56 & 2.30 \\
 $\rho$ & N/A & 0.003 & 0.022 \\ 
 Components & 447 & 447 & 447 \\
 GCC Size & 116,411 (99.2\%) & 86,364 (99.4\%) & 30,047 (98.5\%) \\
 \end{tabular}
 \caption{Bipartite and projection network features}
 \vspace{-10px}
\label{tab:proj}
\end{table}

We omit measuring the clustering coefficient $\mathcal{C}$, degree
assortativity $\mathcal{A}$, and density $\rho$ of bipartite network because their 
definitions are closely related to measurements taken over the network's
{\em one-mode projections}~\cite{guillaume06}.
One-mode projections capture the structure of 
interaction co-occurrences among the $g$ listeners and $n$ members
of 7cot.
Given a matrix $\mathbf{B} \in \mathbb{R}^{g \times n}$ where 
$\mathbf{B}_{ij} = 1$ if listener $i$ has a conversation
with member $j$, we define $\mathbf{P}^{(m)} = \mathbf{B}^T\mathbf{B} \in \mathbb{R}^{n \times n}$ and 
$\mathbf{P}^{(l)} = \mathbf{B}\mathbf{B}^T \in \mathbb{R}^{g \times g}$ as the adjacency matrices 
of the member and listener projection networks, respectively. We then have $\mathbf{P}^{(m)}_{ij} = c$ 
($\mathbf{P}^{(l)}_{ij} = c$)
if members (listeners) $i$ and $j$ hold a conversation 
with $c$ common listeners (members). Structural patterns within the
projection networks are discussed next. 

\begin{figure}
    \subfloat[Member network sample]{\includegraphics[width=1.74in]{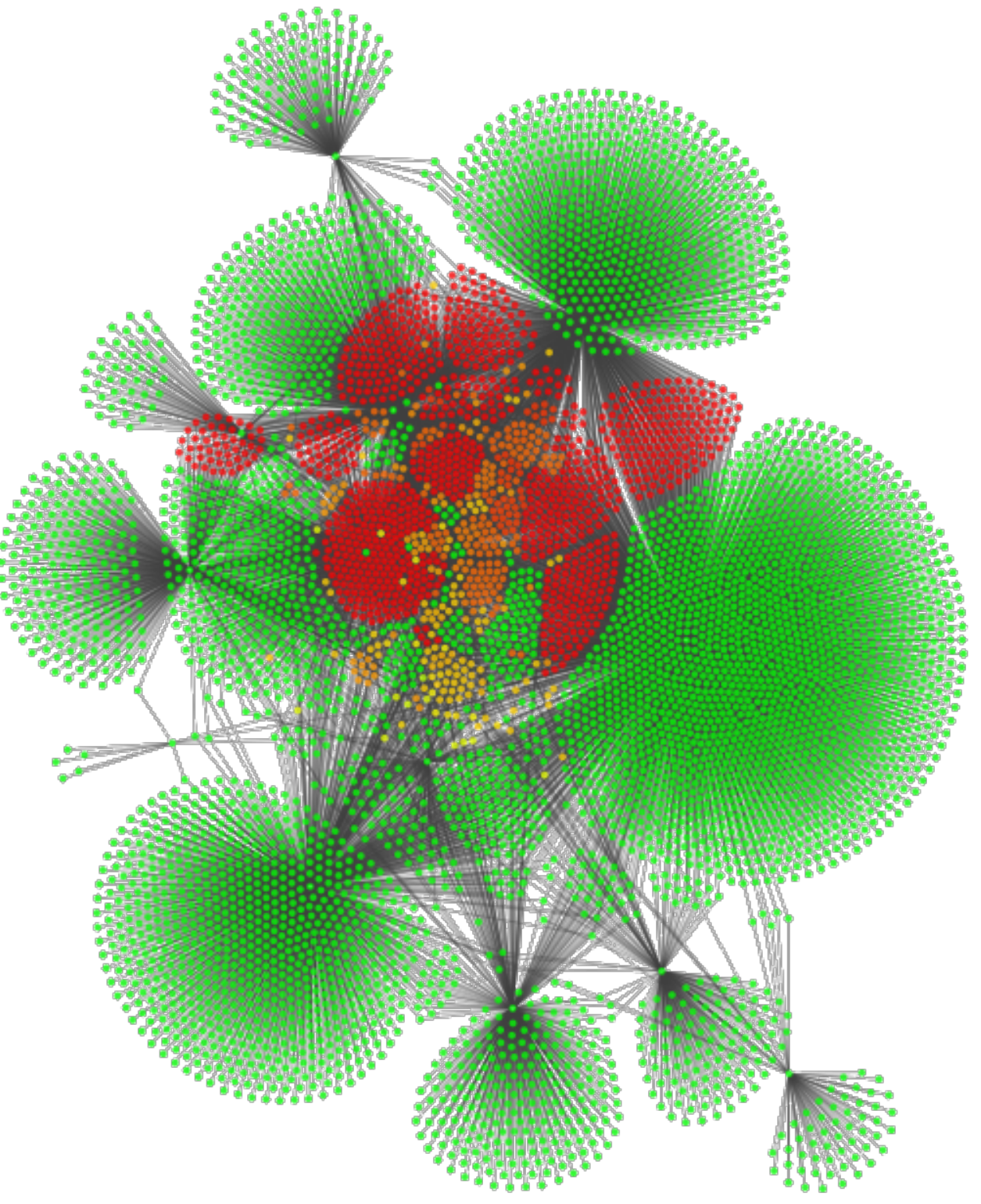}%
  \label{fig:memsample}}
\hfill
\subfloat[Listener network sample]{\includegraphics[width=1.74in]{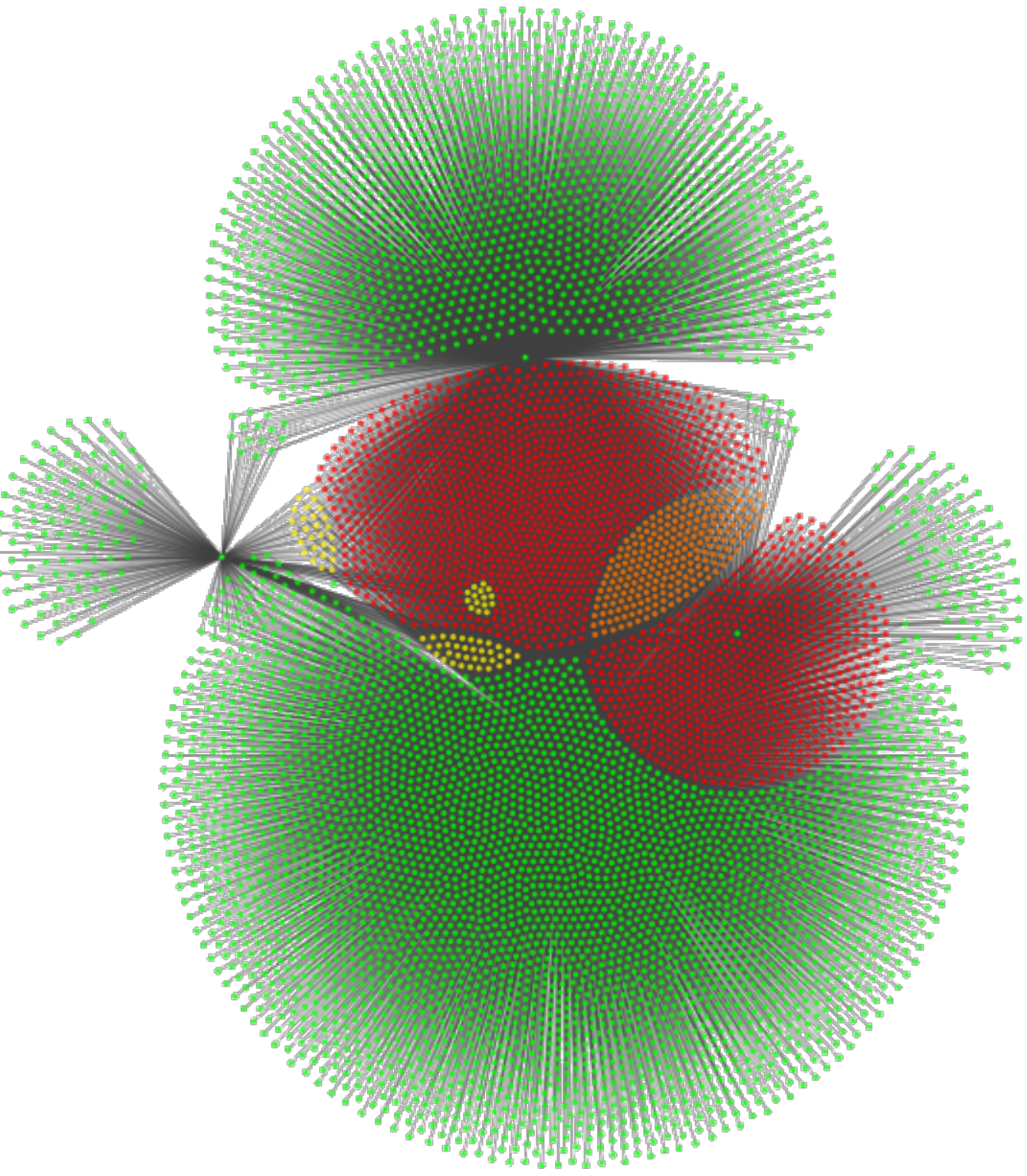}%
  \label{fig:listsample}}
  \caption{Edge sampled projection networks with nodes colored by clustering coefficient}
  \label{fig:sample}
  \vspace{-15px}
\end{figure}

\subsection{Connectivity patterns}
Table~\ref{tab:proj} gives the mean degree, global clustering coefficient, degree assortativity, average path length, 
density, and GCC size of the member and listener projection networks. 
These statistics may be compared with a visualization of a random sampling~\cite{ahmed12} of 
10,000 edges of the projection networks in Figure~\ref{fig:sample}. 
Nodes are colored hotter in the figure if they 
have a higher local clustering coefficient $\mathcal{C}_l$ 
(green nodes have $\mathcal{C}_l = 0$ and red nodes have $\mathcal{C}_l = 1$) 
 and are 
drawn under a force directed layout so that nodes separated by small
distances are positioned closer together. 
Although sophisticated sampling algorithms are needed to create samples that maintain many
structural features of the sampled network~\cite{dyad14}, edge sampling still conveys the shape
of the global network within the interconnected core of the sample
(nodes participating in excessive numbers of open triangles are likely an artifact
of edge sampling).  
The high mean degree, large GCC size, and small average path lengths of both projections further
support the hypothesis that members and listeners do not limit themselves to interact 
with a small subset of listeners (members). They both exhibit weak negative degree 
assortativity, suggesting a small inclination
for members (listeners) who share just a few common listeners (members) with others share
them with those who have large numbers of listeners (members) in common with others.
However, the lower degree, larger clustering
coefficient, and larger path lengths of the member network imply a weak penchant for members 
to form clusters by the common listeners they connect to. Such clusters
can be seen in Figure~\ref{fig:memsample} as cliques in the core of 
the member network. 
These clusters may be traces of member groups that connect to similar `types' of listeners. 
\begin{figure}[t]
\vspace{-25px}
\subfloat[Member projection]{\includegraphics[width=1.74in]{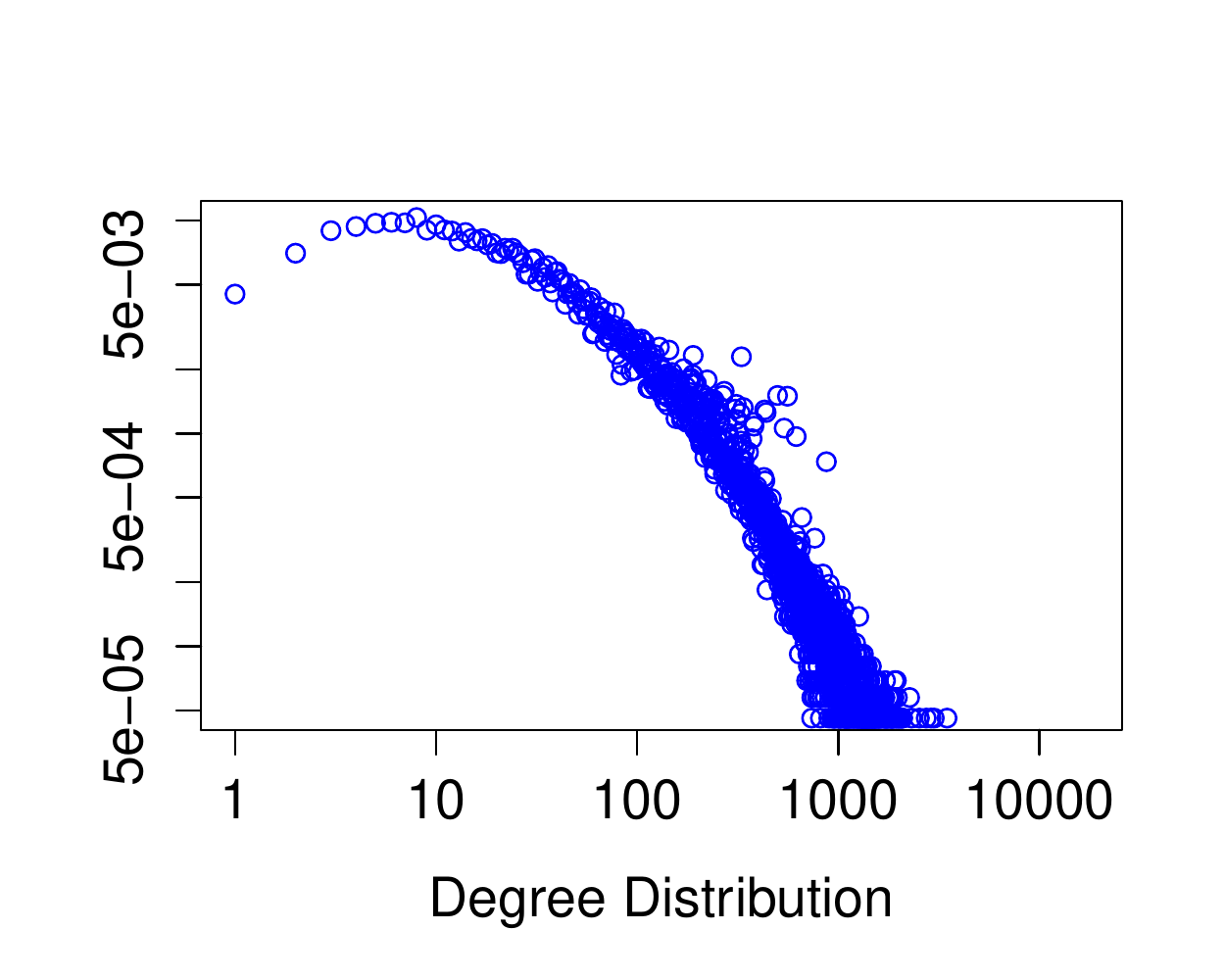}%
  \label{fig:ddmem}}
\hfill
  \subfloat[Listener projection]{\includegraphics[width=1.74in]{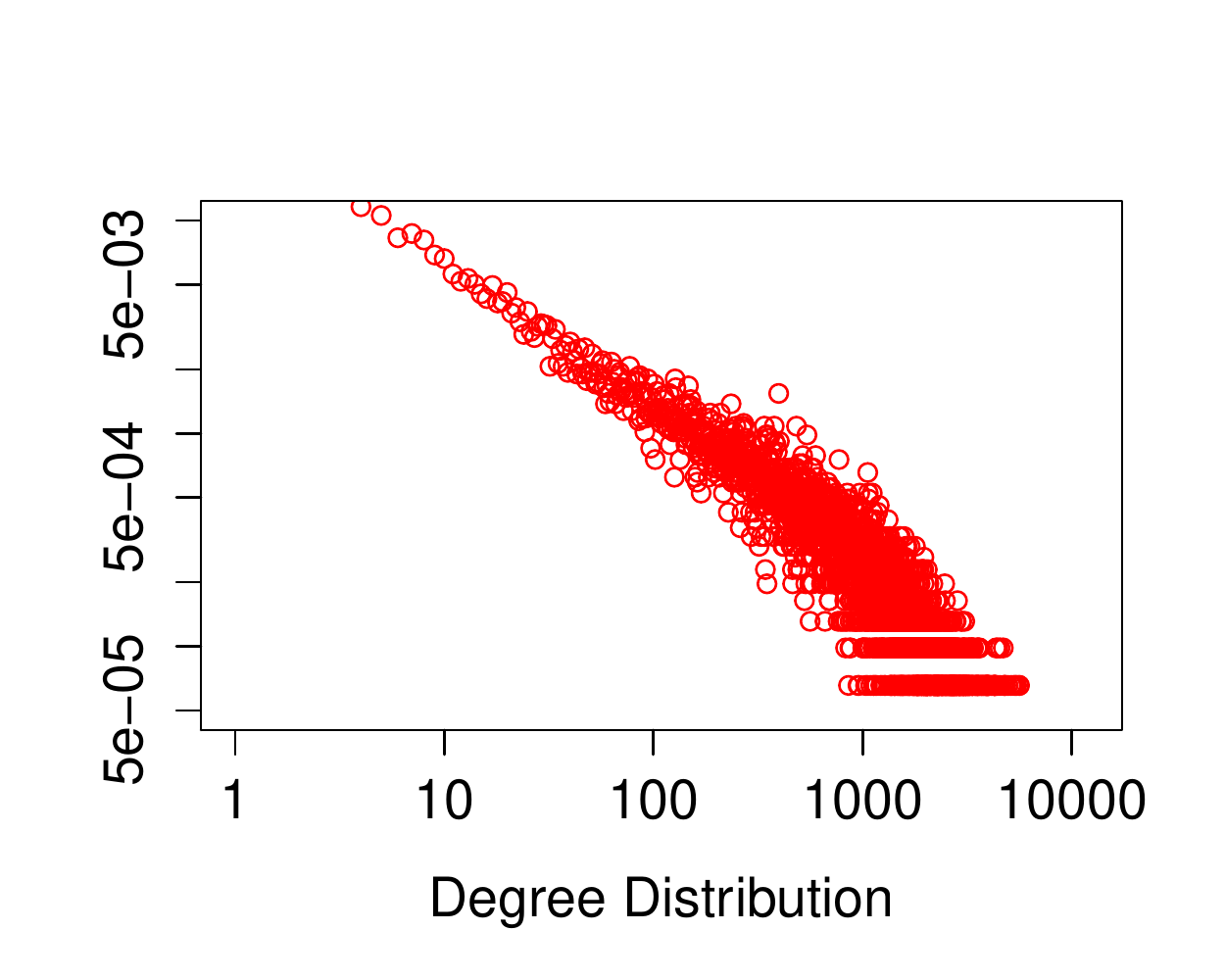}%
  \label{fig:ddlist}}
  \caption{Projection network degree distributions}
  \label{fig:dd}
  \vspace{-15px}
\end{figure}

We find the degree distributions of the projection networks, presented in log-log
scale in Figure~\ref{fig:dd}, to take dissimilar shapes. The listener degree
distribution exhibits a near straight line pattern
indicative of a power-law distribution, but the  pattern is less pronounced 
in the member degree distribution. 
We quantify this difference by running a maximum likelihood based test of the null hypothesis 
{\em $H_0$: the empirical data has a power-tailed distribution} (the test also yields
best fitting power-law exponent $\alpha$ under the null)~\cite{clauset09}.
The test leaves little room to reject $H_0$ for the listener degree distribution 
($p = 0.985; \alpha = 2.51$), but there is more doubt for
the member degree distribution ($p = 0.362; \alpha = 2.34$). 
That the listener degree distribution has a
power-tail suggests significant variation in the number of common members listeners share
with each other, and that the probability of sharing orders of magnitude more members than
expected is not negligible. A similar statement could be made about members, however 
they may exhibit less variation since we are less confident if a power-tailed trend exists. 
The difference of the distributions shape may be explained by members who only need to connect to a limited
number of listeners in order to have many problems resolved, or by members who choose to 
connect deeply with a small number of listeners. Such behaviors place
a `soft limit' on the largest number of listeners members may connect to, weakening the support for a 
power-tail to emerge~\cite{laqt}.  On the other hand, so long as a listener is available for 
newly added members to connect to, there may be no limit on the number of 
new members a listener may connect to over time.

\subsection{Centrality analysis}
\begin{figure}[t]
\vspace{-25px}
 \subfloat[]{\includegraphics[height=1.28in]{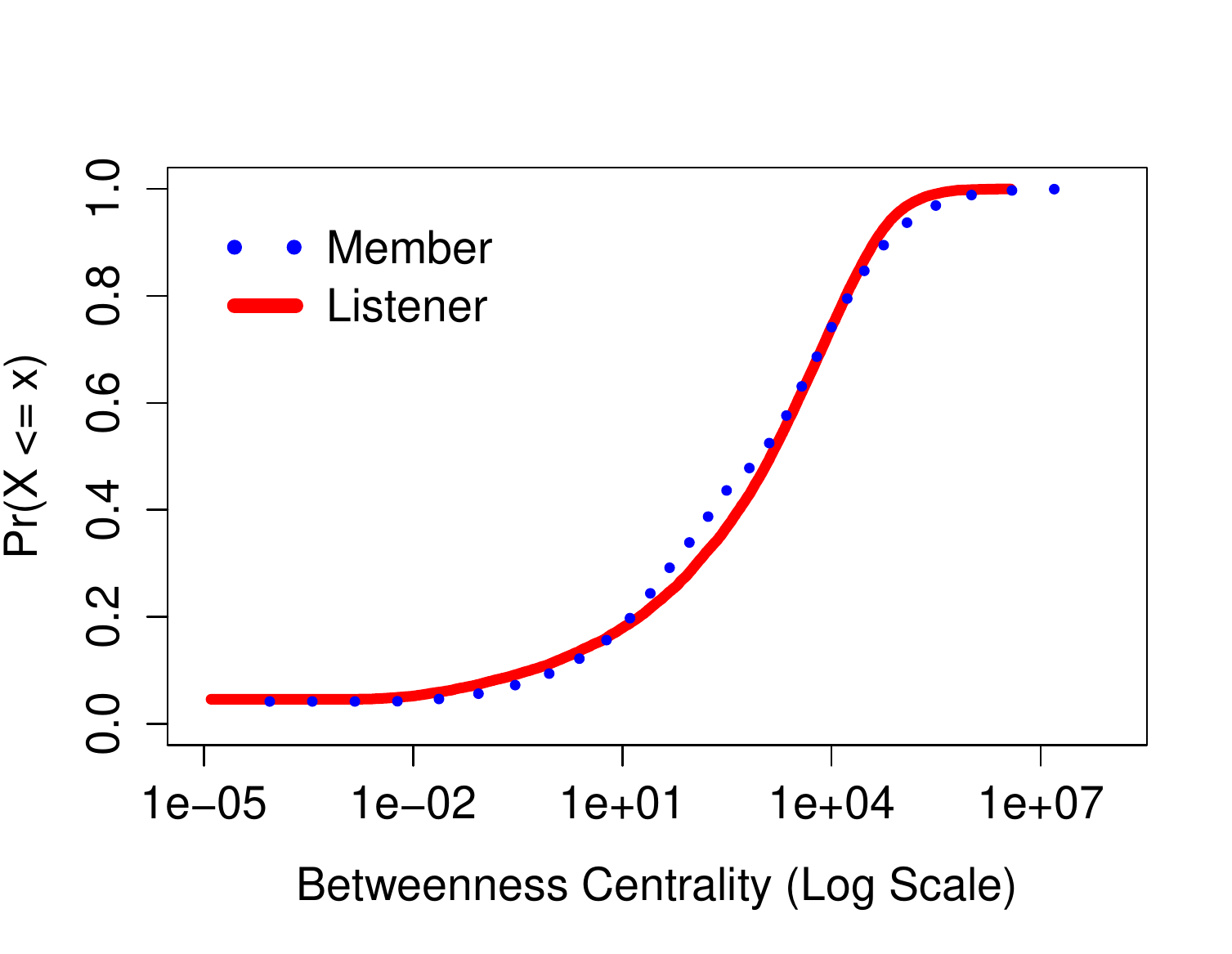}
 \label{fig:btcdf}}%
 \hfill
 \subfloat[]{\includegraphics[height=1.28in]{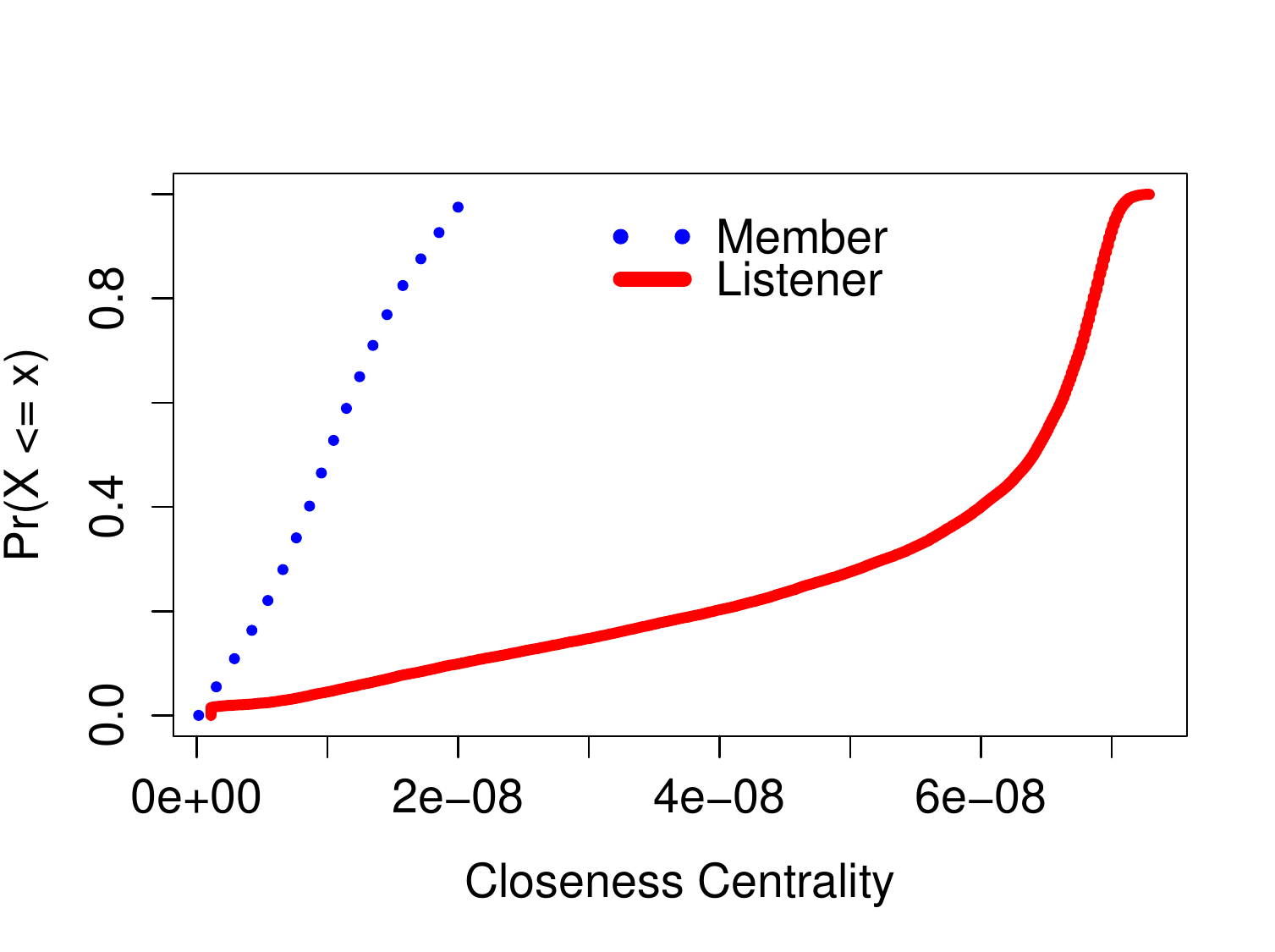}
 \label{fig:clcdf}}%
  \caption{Centrality distributions for members and listeners}
  \vspace{-15px}
\end{figure}

We also study connectivity-based notions of network centrality in the 
projection networks. We first consider the betweenness centrality
of a user $u$, defined as 
$b(u) = \sum_{u\neq i \neq j} \sigma_{ij}(u) / \sigma_{ij}$ where 
$\sigma_{ij}$ is the number of shortest paths from users $i$ to $j$ and
$\sigma_{ij}(u)$ is the number of such paths that include $u$. 
This measure reflects the notion that a user is `central' if she is often part of the 
shortest path among two others in the network. Figure~\ref{fig:btcdf}
plots the cumulative distribution (CDF) of the centrality scores across the 
two networks on semi-log scale. Its rapid ascent and long left tail
indicate that almost all users are part of a number of shortest paths in the network. 
The networks are therefore structurally robust to the loss of users. 
We also consider the closeness centrality of a user $u$,
defined as $c(u) = (\sum_{j} d(u,j))^{-1}$ where $d(u,j)$ is the distance
from user $u$ to $j$. Figure~\ref{fig:clcdf} gives the CDF of closeness centrality
on the two networks (note that the $x$-axis is not in log scale). That the
CDF for the listener distribution is stretched farther than the 
member distribution is only because there are fewer nodes in the network.
Unlike betweenness centrality, the closeness centrality CDF 
of the two networks takes on different shapes. 
The CDF of the member network has only a slight
curvature at its left and right tail, with a nearly linear body. This suggests
that the centrality scores exhibit a small peak around
the mean of the distribution but are otherwise uniformly distributed.
The centrality scores of listeners are uniformly distributed up to 
approximately the $40^{th}$ percentile, at which point they become
heavily skewed. A majority of listeners, therefore, 
are at a much shorter distance from those below this $40^{th}$ percentile.
This pattern may be indicative of a core-periphery structure~\cite{rombach14}
in the listener projection network that does not exist in the member one, 
where those in the core (periphery) have high (low) closeness centrality. 
The probability of a listener falling in the core may be correlated with the 
diversity of the members she connects to: connecting to many different
members increases the probability of sharing a connection with a listener
already in the core. 

\subsection{Network transitivity}
\begin{figure}[t]
\vspace{-0px}
   \subfloat[Member projection]{\includegraphics[width=1.74in]{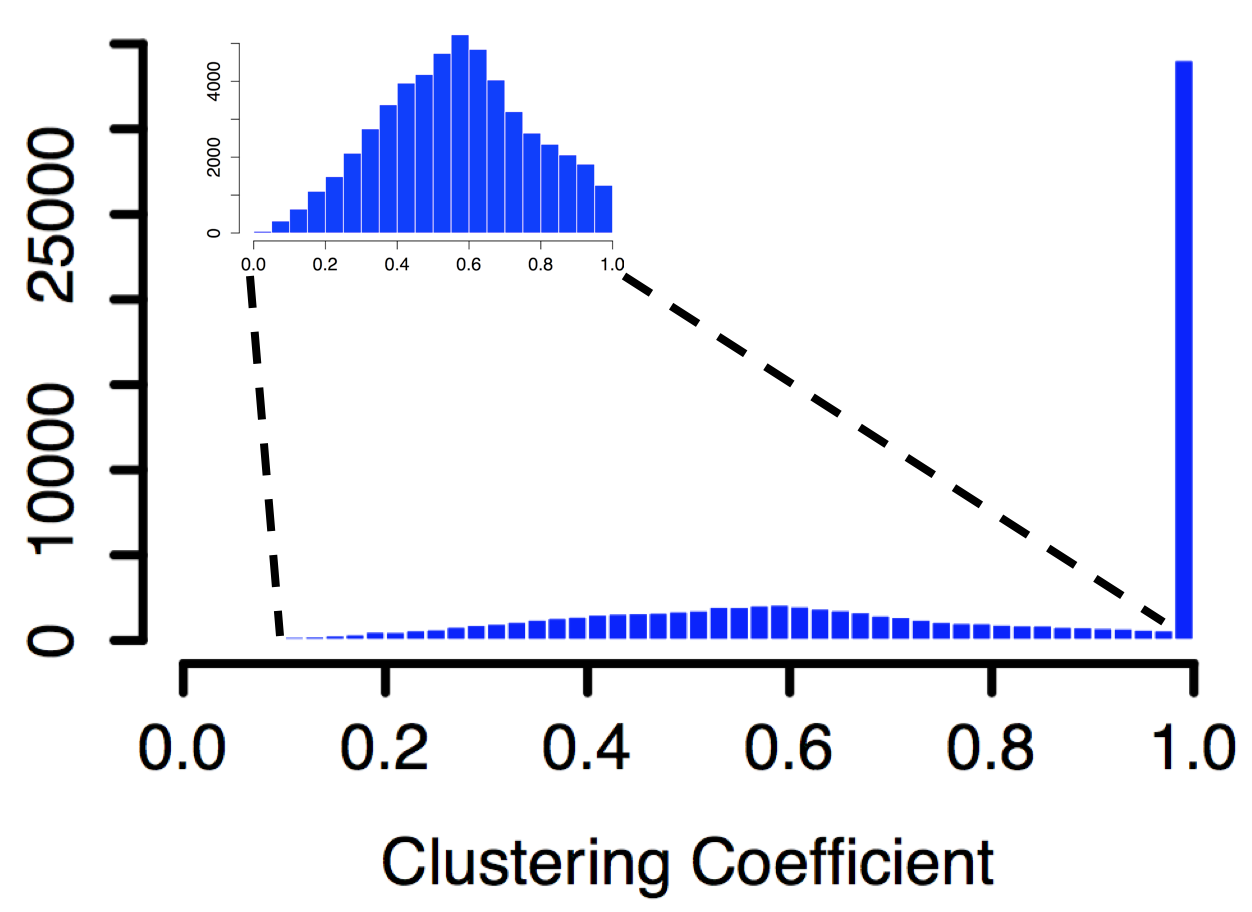}}%
 \hfill
   \subfloat[Listener projection]{\includegraphics[width=1.74in]{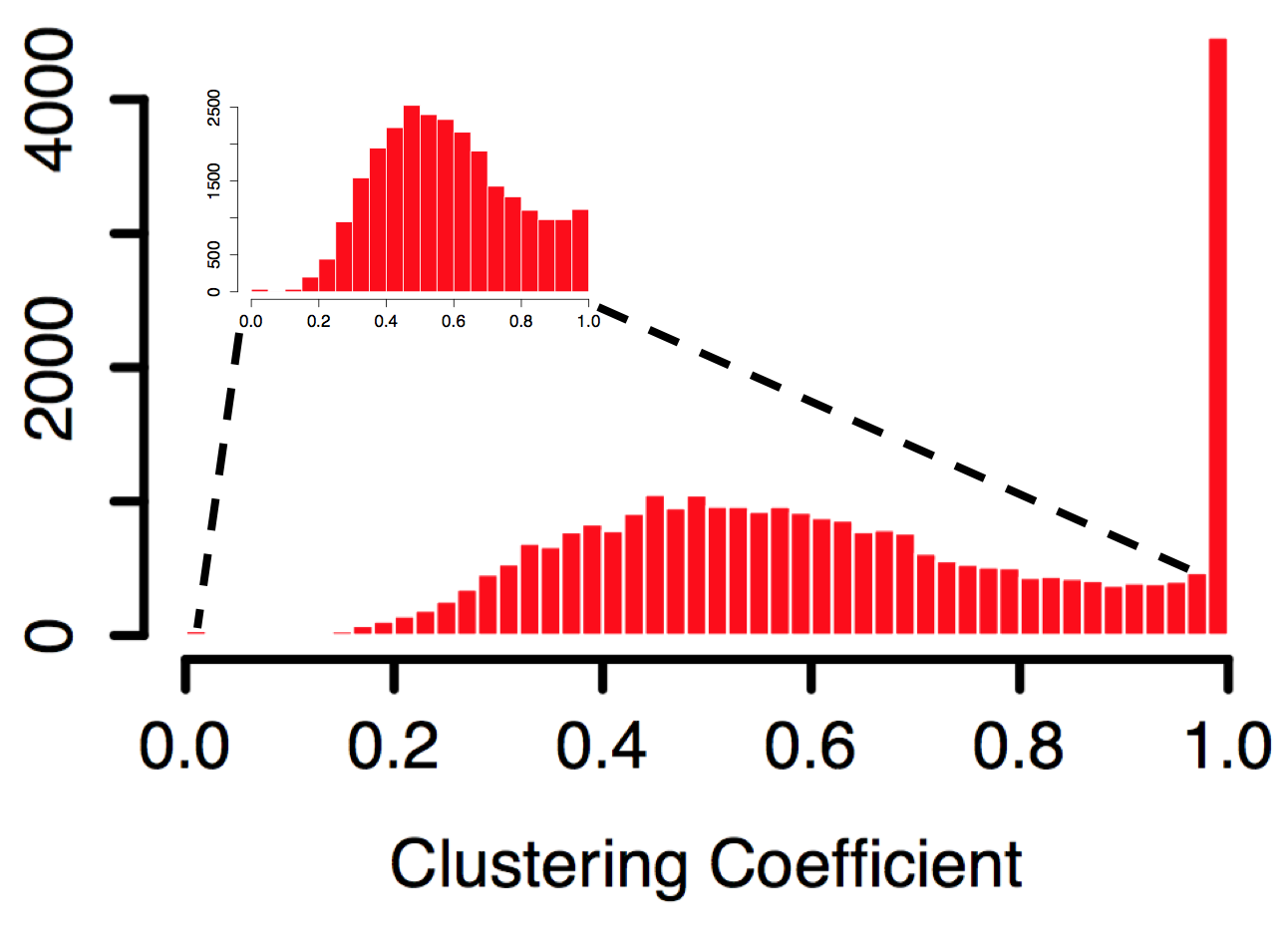}}%
  \caption{Projection network cluster coefficient distributions}
  \label{fig:cc}
  \vspace{-15px}
\end{figure}

Finally, we use the local clustering coefficient distributions of the projection 
networks to study the tendency of transitive relationships among members
and listeners. A transitive relationship is one where if user $A$ is a 
member (listener) connected to user $B$ and $B$ is connected to $C$, then $A$
is connected to $C$. Table~\ref{tab:proj} lists the global clustering coefficient,
defined as the average of the number of closed triangles in a user's neighborhood 
divided by the number of possible links that could exist within 
it~\cite{watts98}, as $\mathcal{C} = 0.734$ and $0.636$
for the member and listener projections respectively. The large  
coefficients signify that transitive relationships dominate the projection
networks. However, histograms of the local clustering coefficients $\mathcal{C}_l$ in the
member and listener network in Figure~\ref{fig:cc} show that the large values are driven 
by the 38.9\% of members and 13.2\% of listeners whose $\mathcal{C}_l = 1$. The high values
of $\mathcal{C}$ are therefore driven by a small proportion of users with fully
connected neighbors. 
When we consider users whose $\mathcal{C}_l < 1$, 
closeness centralities appear to be normally distributed.  
Normally distributed $\mathcal{C}_l$ distributions is a typical phenomenon in co-occurrence
networks spanning many systems, including scientific paper authorship~\cite{yang15,leskovec07g}, 
e-commerce co-purchases~\cite{leskovec07d}, and ``related page" relationships on
search engines~\cite{niu11}, but the surge of members where $\mathcal{C}_l = 1$ is 
unique to 7cot interactions. 
This suggests that users with $\mathcal{C}_l = 1$ may not emerge from some natural or 
universal process innate to all co-occurrence networks. This is evidence that 
both members and listeners perform deliberate actions that drive them into fully connected 
neighborhoods in the projection networks. For example, members may be selectively
connecting to the same pool of listeners that may have similar ratings, 
experiences, or bio's suggesting an expertise that members in their neighborhood do.
Finally, it is interesting to note that the proportion of members where $\mathcal{C}_l = 1$ (38.9\%)
is very similar to the proportion of personal conversations (where a member chooses
a listener to connect to) in Table~\ref{tab:summary} (39.8\%).

\section{Understanding User Engagement}
\label{sec:engagement}
Next, we perform an engagement analysis of members on 7cot. 
Engagement analysis offers insights about the user and platform features 
that encourage members to return, listeners to stay active, and encourage members to have multiple,
fruitful conversations. Such insights are practically important to 
help a platform retain new members and grow its community of listeners. 
They also identify qualities that encourage people to seek 
follow-up emotional support. Due to space limitations, we will consider
listener engagement in future work. 

\begin{table}
\centering
\begin{tabular}{r | c || r | c}
Coins & 0.247 & Growth Points & 0.977\\
Compassion Hearts & 0.243 & Signup Date & -0.009 \\
Last Login Date & 0.133 & Distress Level & 0.004 \\
Group Chat Msgs & 0.120 & Page Views (Web) & -0.002\\
Page View (iOS) & -0.001 & Login Count & -0.001 \\
Conv. Requests & 0.001 & Self Help Views & 0.005\\
Forum Posts & -0.001 & Forum Views & -0.001\\
Forum Up-votes & 0.201 \\
\end{tabular}
\caption{Pearson correlation between message rate and user or behavior features} 
\vspace{-15px}
\label{tab:cor}
\end{table}

\subsection{Factors driving engagement}
We first relate the features and behaviors of members and their relationship
to a measure of site engagement. Since sharing with listeners is the purpose of the service, we quantify engagement as the 
{\em message rate} of a 
member, that is, the average number of messages sent per day in conversations.
We consider features and behaviors that, based on discussions with psychologists and designers at 7cot, may be related to 
engagement:~(i) 
number of coins, growth points, and compassion hearts, which are gaming
and progress measures related to a members reputation and 
experience;~(ii) signup and last login date;~(iii) reported distress level 
when members register;~(iv) 
number of group chat messages;~(v) number of page views from the 7cot Web and iOS 
applications;~(vi) number of logins;~(vii) number of conversation requests 
sent;~(viii) number of self help page views;~(ix) number of forum posts, views, and up-votes.
Table~\ref{tab:cor} gives the Pearson correlation coefficient between the features and a
members' message rates. The coefficients make clear that the gamification 
features of the platform (accumulated coins, hearts, and growth points) are 
strongly related to the engagement of a member. However, conversation messages sent
by members directly increase growth points, giving this correlation little
meaning. Member attributes
and behaviors unrelated to communication (signup and last
 login date, distress level, page views, and help article views) exhibit virtually
 no correlation, suggesting that users dealing with any type and
 degree of emotional distress, at any time, 
 exhibit similar levels of engagement on the site. 

Many features exhibit little correlation 
with user engagement, but interaction terms built by subsets of them may be 
positively correlated. 
For example, users who exhibit a high distress level and submit many conversation
requests may have a high level of engagement even though the features are 
individually not correlated. Instead of exhaustively
exploring all multi-way interactions, we consider
a random forest model that predicts user engagement by a regression over 
all features. 
A random forest is an ensemble of decision trees, each of which is trained over
different bootstraps of the data. During training, each tree is limited to the use
of distinct small subsets of the features to make splitting decisions. 
If $X_u$ is a vector of member $u$'s features, the random forest predicts the 
engagement of $u$ as $\hat{f}(X_u) = N^{-1}\sum_{i=1}^N \hat{f}_i(X_u)$
where $\hat{f}_i$ is the predicted engagement value from the 
$i^{th}$ of $N$ decision trees in the random forest. The bootstraps,  
limited choice of features for tree splitting, and averaging of results
across the tree ensemble ensure the forest does not overfit the data even for 
large $N$~\cite{hastie09}. We compute the importance of each feature to the 
random forest regression model as follows: let
$C = \sum_{i=1}^m (y_i - \hat{f}(X_i))^2$
be the mean square error (MSE) of the random forest predictions against 
the actual engagement $y_i$ of every member $i$. The importance
of feature $\ell$ may be found by randomly perturbing the values of $\ell$ 
across every member's feature vector. Letting $X^{(\ell)}_i$ 
be the feature vector of member $i$ whose $\ell^{th}$ element is perturbed and 
 $C_\ell = \sum_{i=1}^m (y_i - \hat{f}(X^{(\ell)}_i))^2$ as the MSE
of the model using the perturbed vectors, the importance of $\ell$ may ranked 
by the percent increase in MSE between $C$ and $C_\ell$. For example,
if feature $\ell$ is not important, the errors of the model 
will be less sensitive to a reshuffling of its values across all users. 

\begin{figure}
\vspace{-15px}
	\subfloat[Quantiles]{\includegraphics[width=1.73in]{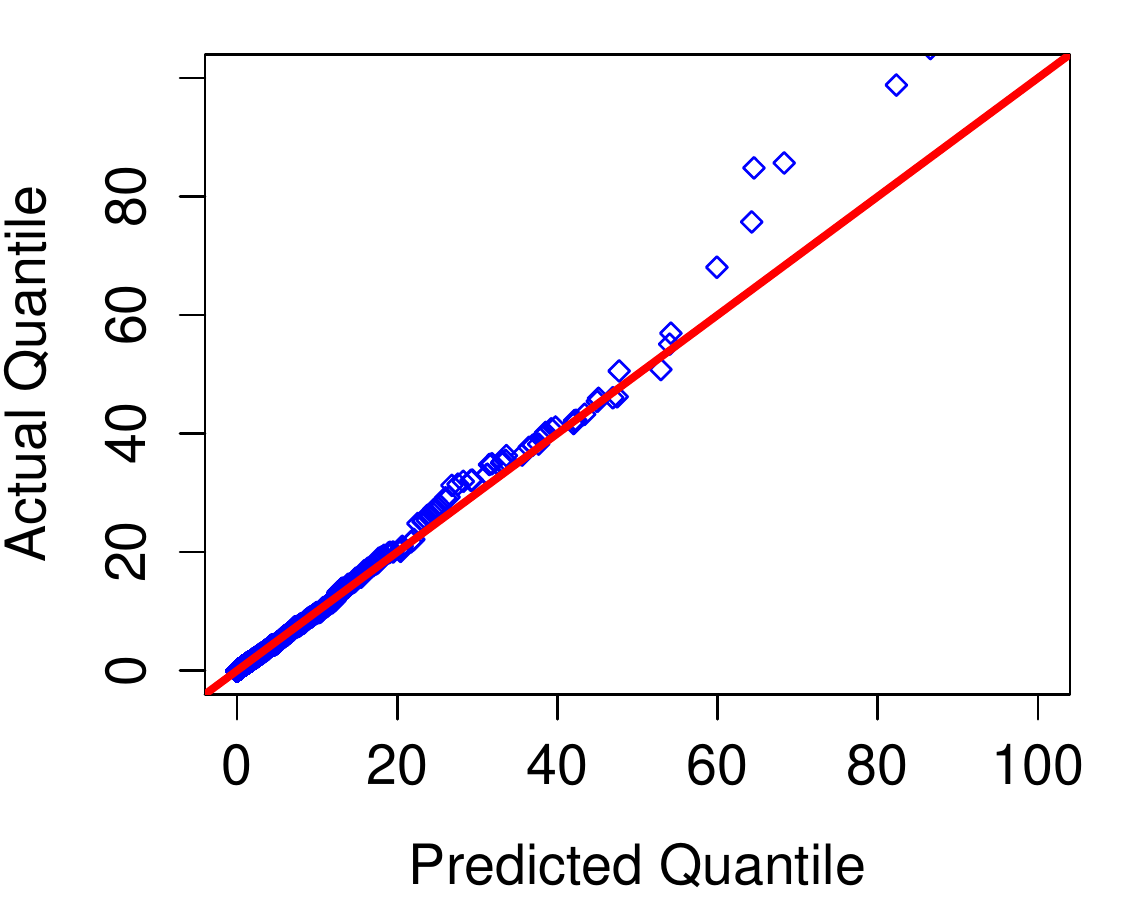}%
	\label{fig:forest_qq}}
\hfill
	\subfloat[Predictions]{\includegraphics[width=1.73in]{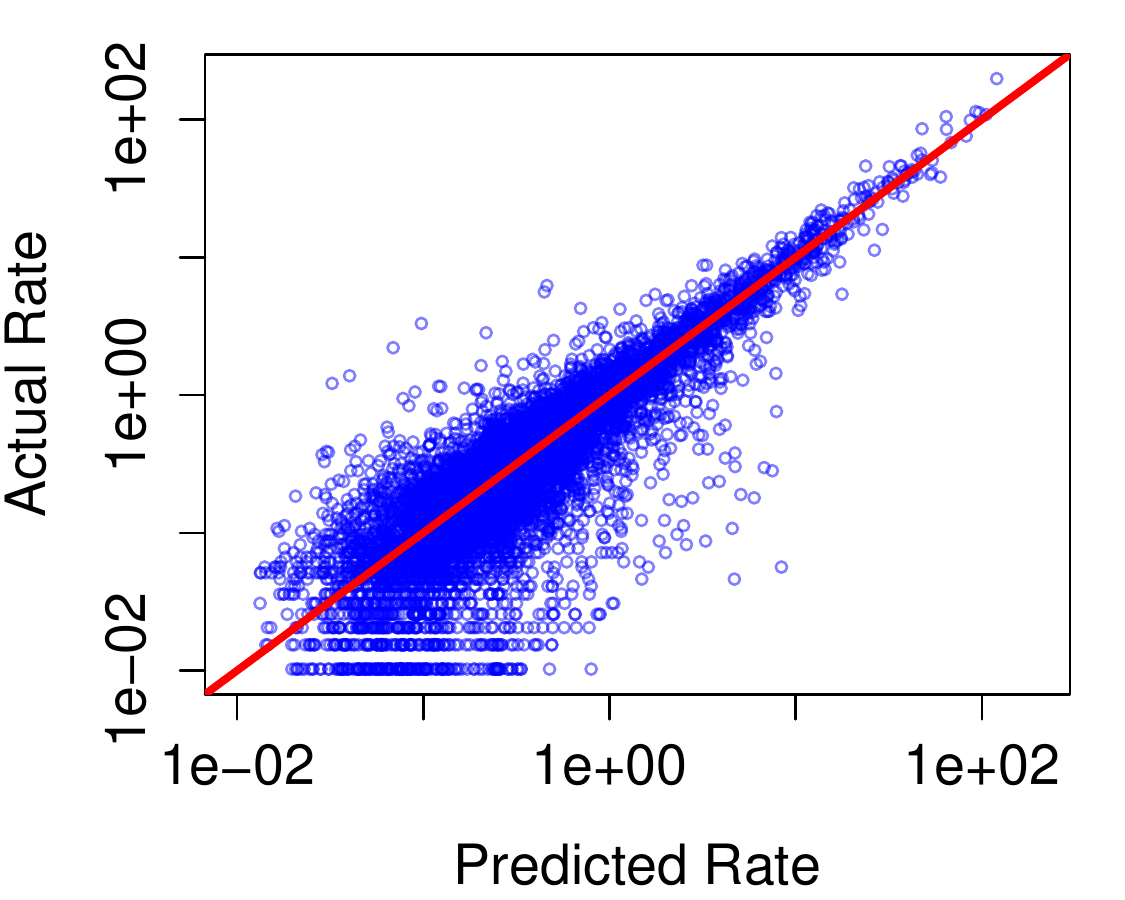}%
	\label{fig:forest_predict}}
\caption{Random forest predicted vs. actual engagement}
\label{fig:forest}
\vspace{-15px}
\end{figure}

We trained a random forest using 
75\% of the user data for a forest 
with $N=1000$ trees and randomly choose 1/3 of the features
for every tree splitting decision. 
Figure~\ref{fig:forest} gives the quantile and prediction
scatter plots of the predicted and actual message rates 
for the 25\% of users not used to train the random forest. 
The figure demonstrates that the decision tree models
engagement very well ($R^2 > 89\%$), as the quantile plot shows a 
linear relationship between the distribution of the predicted and actual
engagement rates up to the $60^{th}$ quantile. 
The predicted vs. actual engagement rates 
in Figure~\ref{fig:forest_predict} only
show normally distributed errors for users with low engagement. 

\begin{figure}
\centering
  \includegraphics[width=2.5in]{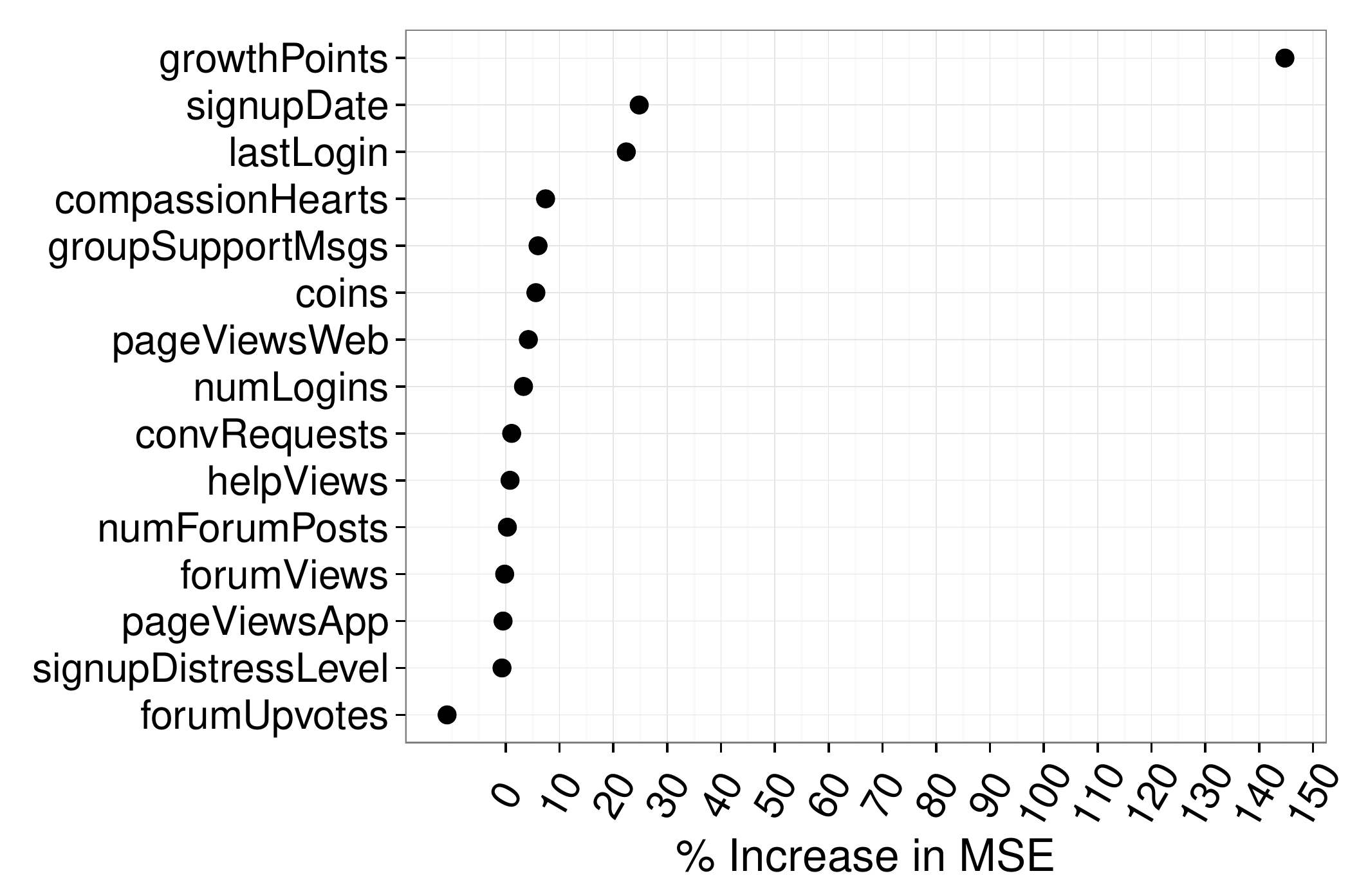}%
  \caption{Feature importance in the random forest model}
  \label{fig:mse}
  \vspace{-15px}
\end{figure}

Since the random forest reasonably
models the relationship between member and behavioral features, 
we use it for feature importance analysis. 
Figure~\ref{fig:mse} shows the percent increase in MSE of a random forest
trained with data where each factor was individually perturbed across the training data. As
anticipated by Table~\ref{tab:cor}, the total number of growth points of
a member is the most important factor for predicting user engagement due to its 
direct correspondence with her message rate. Members' signup and last login
dates are the next most important features, each of which increases MSE by 
over 20\% when they are perturbed. The signup date of a user is weakly
anticorrelated with engagement according to Table~\ref{tab:cor}, thus recent 
logins have a weak relationship to engagement. 
The number of messages sent in group chat is the next non-gaming related
feature that is important for user engagement. 
This suggests that participating in group chats encourages users to become more
engaged in their one-on-one conversations. It may be the case that users find
group settings to be easier or less intimidating to participate in, and builds their
confidence to have lengthy sessions with a listener. Finally, 
we note that the number of up-votes a member has on the forum actually introduces
noise in the model, since perturbing this factor decreases the MSE of 
the random forest. One explanation may be that members who gain recognition
for their forum posts may be disinclined to participate in 
conversations since they achieve recognition and perhaps satisfaction by only
participating on the forum.

\subsection{New user engagement prediction}
New members to a service may be active for a brief period of time,
then become `inactive' and never return. Early identification of new
users likely to become inactive helps a platform identify those who could be 
encouraged or incentivized to continue seeking help, or become listeners to bolster its community. 
Feature importance analysis of such an identifier may also reveal
the behaviors and attributes that promote people to return and seek 
follow-up support.

We consider a random forest classifier that identifies if a member, based on
actions during her first two weeks on 7cot, will become an active user. 
Since there is lack of a standard definition for an `active' user of a Web service,
we consulted with 7cot administrators to define an active user as one who:~(i) has been
registered for at least six weeks; and~(ii) has performed at least two actions 
on the service over the past month. We also define a `new user' as one who has registered 
within the last two weeks. We identified all members who registered
between May 7th (the first date user action data was recorded) and November 18th, 2014 (the end of our data set) 
and mark them as `active' or `inactive'. We then collected the following actions they performed
during their first two weeks on the site:~(i) number of conversation requests and 
messages sent;~(ii) number of forum posts made and viewed;~(iii) number of logins performed;
~(iv) number of help page views; and~(v) number of site pages accessed via 7cot's Website and 
iOS app.

\begin{figure}
\vspace{-15px}
  \subfloat[ROC curve]{\includegraphics[width=1.745in]{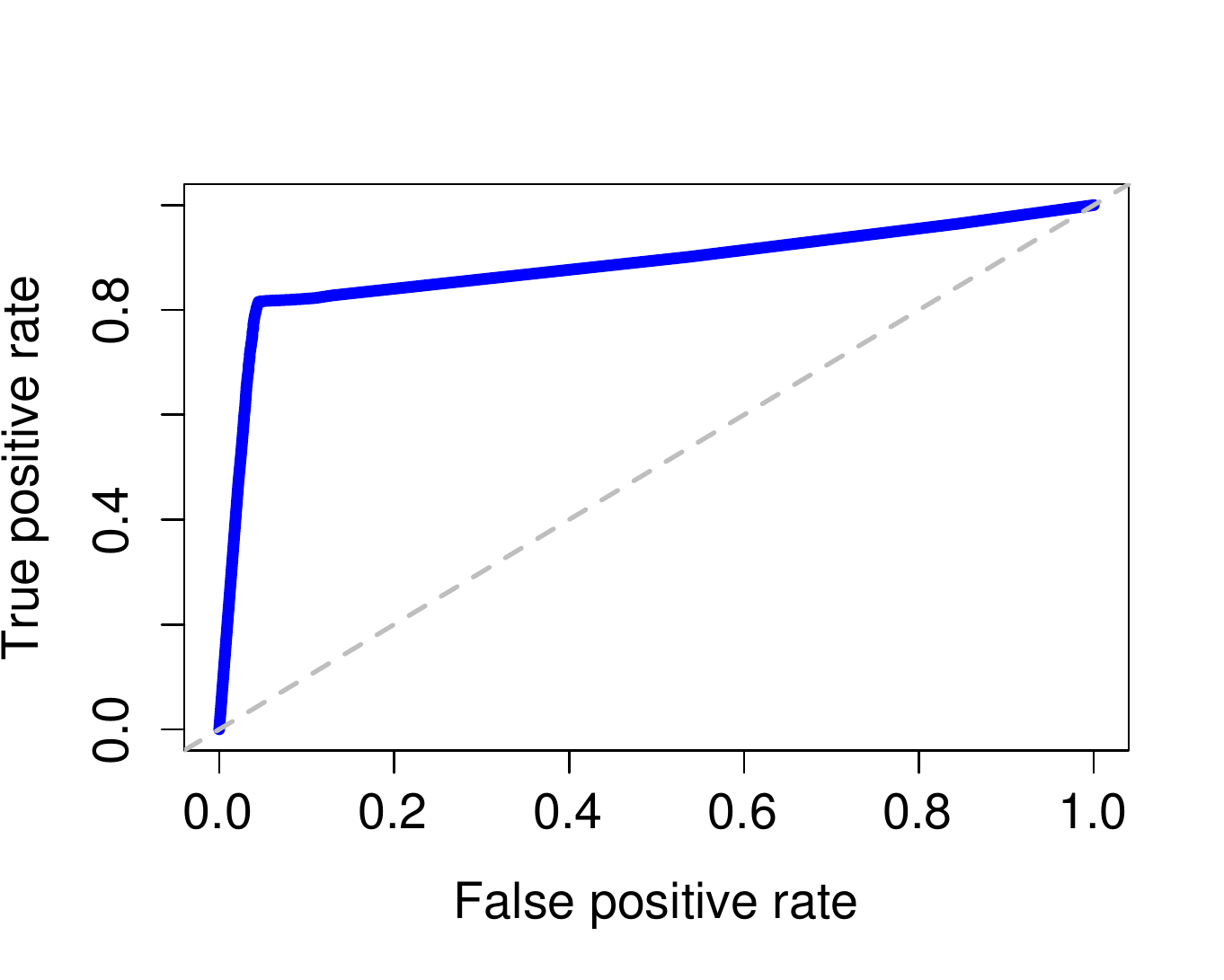}%
  \label{fig:roc}}
\hfill
  \subfloat[Gini Coefficient]{\includegraphics[width=1.735in]{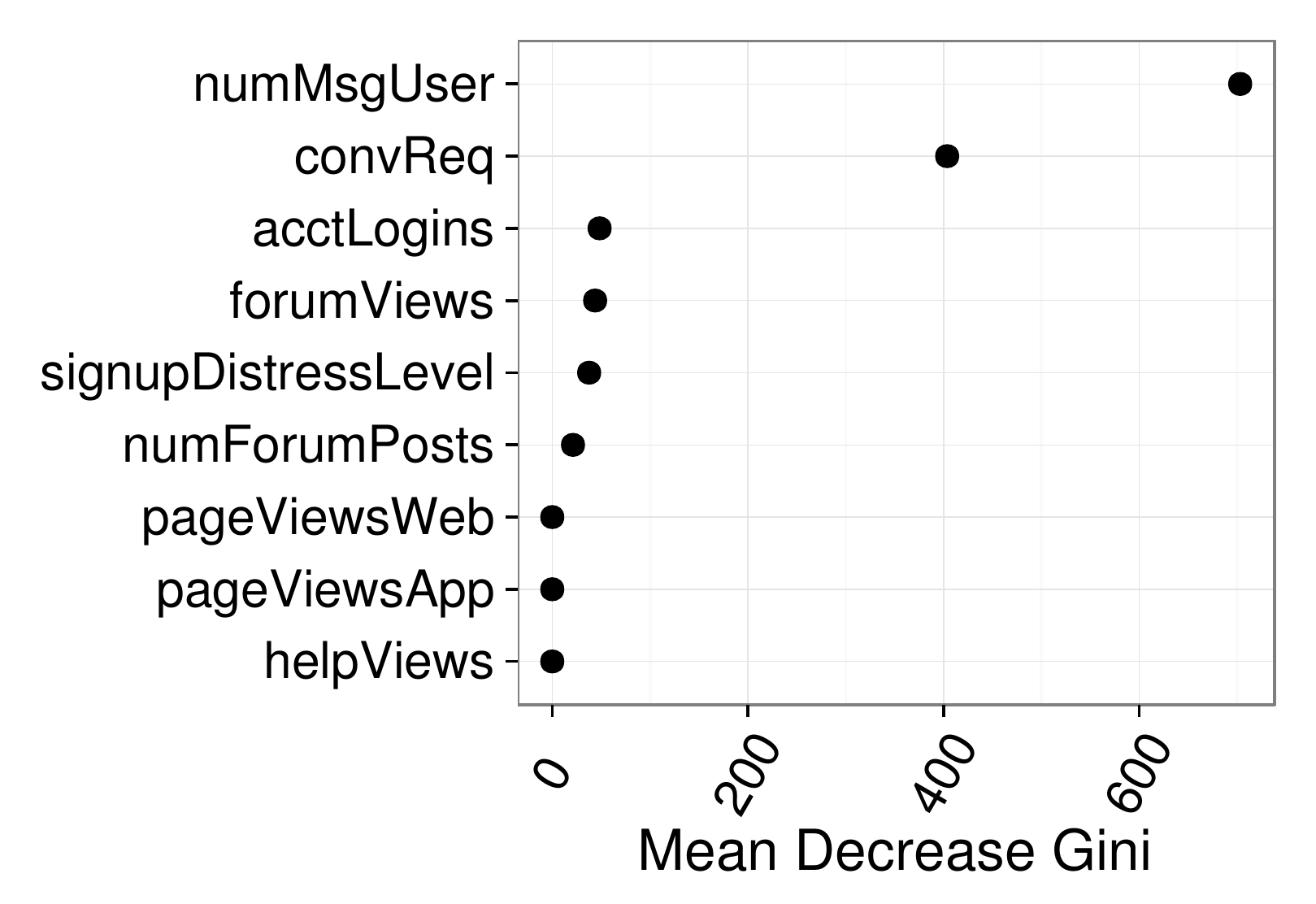}%
  \label{fig:gini}}
\caption{Active user classifier evaluation}
\label{fig:classification}
\vspace{-15px}
\end{figure}

52,803 members registered on 7cot during the time period considered, of which 11,117 (21\%) 
became active and 41,686 (79\%) became inactive. We created a training set by randomly sampling 
66\% of the registered members for a random forest classifier to predict if they are active. 
Trees are trained in a similar fashion to regression. Each tree yields 
its own prediction of if a member will be active or inactive given her 
actions during the first two weeks. A majority vote of the trees then decides the
class to be predicted. Due to the imbalance in the number of inactive and 
active members in the training data, the minority class is randomly oversampled so that 
equal number of inactive and active 
cases are provided for training~\cite{hastie09}. 
The trained random forest was tested over the 33\% of users not
considered in the training set. The classifier achieves a very promising accuracy of 
92.5\% and the ROC curve
in Figure~\ref{fig:roc} demonstrates only a moderate false positive rate
(ROC curves approaching the (0,1) corner of the plot are perfect classifiers; the
$y=x$ line represents a classifier that performs random guessing).

\begin{figure}
\centering
\vspace{-0px}
\includegraphics[scale=0.35]{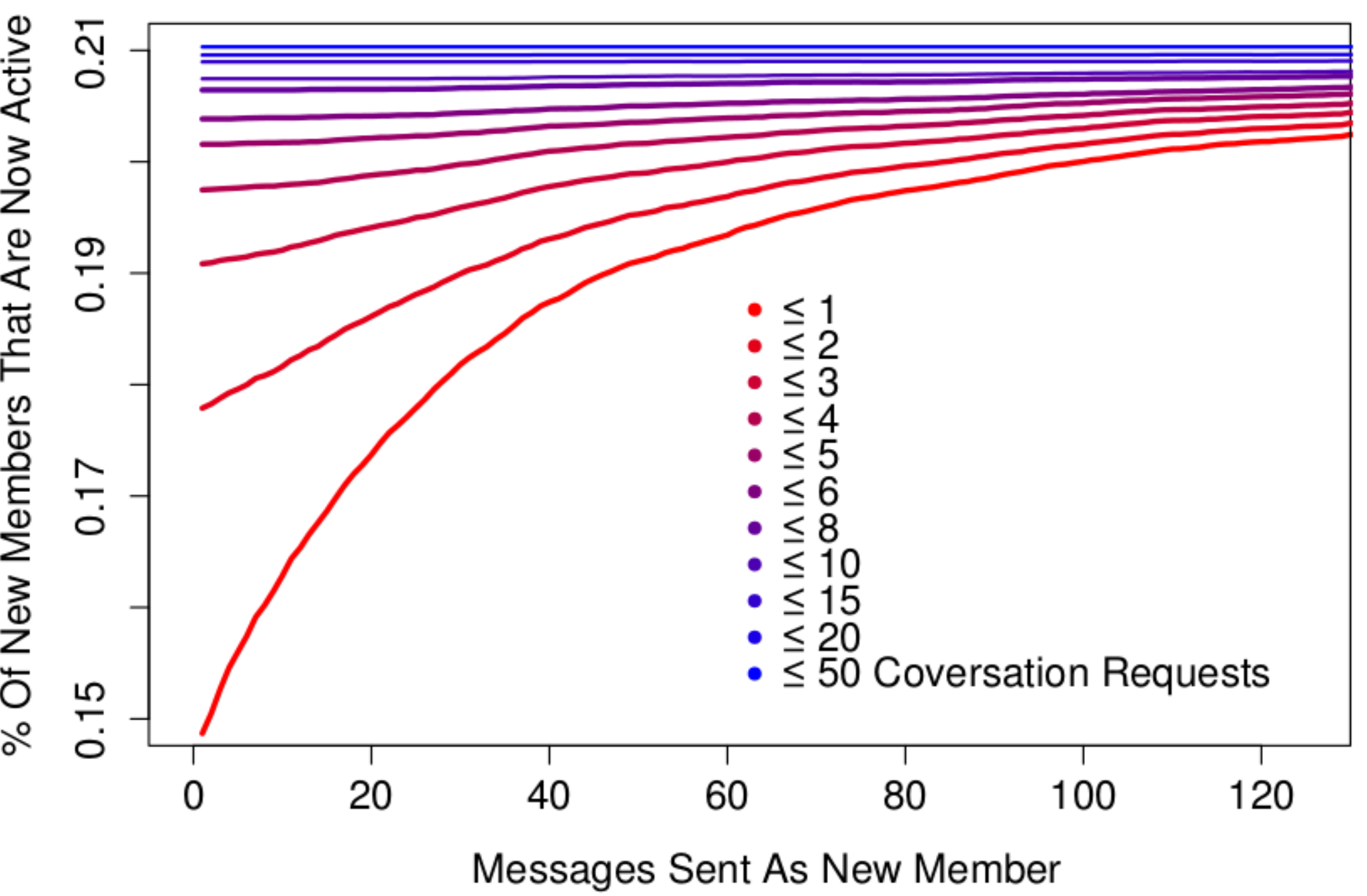}
\caption{Active users retained by number of conversation requests and minimum number of messages sent}
\label{fig:active}
\vspace{-15px}
\end{figure}

As before, we assess the importance of the factors used for predicting active users. 
Since the concept of MSE is incompatible with the notion of a binary classification decision, 
we instead consider the Gini index~\cite{hastie09} of decision tree nodes in the forest. 
The Gini index of a decision tree node $t$ is defined by 
$G_t = p_{ta}(1-p_{ta}) + p_{ti}(1-p_{ti})$ where $p_{ta} (p_{ti})$ is the proportion of members
marked active (inactive) 
that fall into node $t$ based on the splitting criteria of its parent node. A $G_t$ 
close to zero suggests that the splitting rule at the parent divides the data into 
separate classes, which is a property of strong decision tree classifiers. 
We thus rank the importance of a factor by the average decrease of the 
Gini coefficient across all splits in all trees of the forest
in Figure~\ref{fig:gini}. It reveals that the 
number of messages sent in conversations and conversation requests submitted 
within the first two weeks are the actions that best predict whether a user will become active. We further 
examine the interaction between these two features 
by showing the percentage of new users who became active 
and submitted $\leq x$ messages
in their first two weeks in Figure~\ref{fig:active}. 
Each trend corresponds to subsets of members who also submitted less than 
the specified number of conversation requests. The figure
shows how for small numbers of conversation requests, the total number of messages 
sent in one-on-one conversations strongly influences members to become active. 
But once approximately five conversations are created, the 
number of messages sent in a conversation loses its importance. This may be because new users
who connect with greater numbers of listeners feel more obligated to return to these connections
again in the future. On the other hand, when a user connects to only a few listeners, 
a stronger bond between them (i.e., more messages shared) is necessary to drive the member
to return to the site. 

Figure~\ref{fig:gini} also shows that the number of account logins performed, the user's distress level, 
and activity related to the online forums within a member's first two weeks are not major
predictors of her becoming active. 
The frequency with which a member accesses the platform 
is thus unrelated to whether she will become an active member; what matters is not the number of 
times a member visits, but the quality or productivity of those visits as measured by the number
of messages sent and conversations requested. Furthermore, since members are equally likely to 
become active no matter their distress level, people suffering from both basic and complex problems may 
be equally willing to become active in online emotional support platforms. Finally, public spaces to post messages, such as forums, do not
encourage new members to become active ones. This may be because forums serve as a less personal, more public medium of communication.

\section{Conclusions and Future Work}
\label{sec:conc}
As society becomes more reliant on Internet-based communication, online platforms that 
offer emotional support will only grow in importance. This paper presented a detailed analysis
of user interactions on 7cot, which is the largest such platform available today. 
The analysis made important insights relevant to the understanding of an emotional
support platform that could inform the design of future ones.
It shows how users are respectful to each other and tend to use the platform during midweek
evenings, and that the ability for users to access large numbers of listeners 
is important and useful. Structural analysis revealed a small tendency for members to connect
with sets of listeners who have latent common attributes, and that the 
mechanisms letting members connect to any available listener 
may lead to a core-periphery structure of listener relationships. Engagement analysis 
emphasizes the importance of mechanisms to track user progress. In addition, less intimidating
group chats may serve as a gateway to engaging listener conversations, and
users facing both simple and complex problems are as likely to become active
participants. 

Future work will characterize guest attributes and behaviors
 in more detail, and explore how users transition from being guests, 
to members or listeners, and to hybrid users.
The structural analysis will be extended to study the nature of the 
cliques emerging in the projection
networks. This direction 
could reveal popular types of listeners that members search for, 
and whether or not the ailments of members can be inferred based on clique memberships. 
We will also perform a thorough engagement analysis on listeners, to understand
the mechanisms and platform designs that keep them active. Improving the false
positive rate of the active user classifier, alternate quantifications of `engagement', 
and alternative classifier types will also be explored to enhance the engagement 
analysis. 
\bibliographystyle{IEEEtran}
\bibliography{asonam15}

\end{document}